\documentclass{elsart}
\usepackage{graphicx}
\usepackage{epsfig}
\usepackage{amssymb}

\begin{document}
\begin{frontmatter}
\title {Measuring Fast Neutrons with Large Liquid Scintillation Detector 
		for Ultra-low Background Experiments}
\author[usd,ctgu]{C. Zhang}, 
\author[usd]{D.-M. Mei\corauthref{cor}},
\corauth[cor]{Corresponding author.}
\ead{dongming.mei@usd.edu}
\author[usd]{P. Davis},
\author[usd]{B. Woltman},
\author[regis]{F. Gray}

\address[usd]{Department of Physics, The University of South Dakota,Vermillion, South Dakota 57069}
\address[ctgu]{College of Sciences, China Three Gorges University, Yichang 443002, China}
\address[regis]{Department of Physics and Computational Science, Regis University, Denver, Colorado 80221}
\begin{abstract}
We developed a 12-liter volume neutron 
detector filled with the liquid scintillator EJ301 that measures neutrons in an underground laboratory where
dark matter 
and neutrino experiments are located. 
The detector target is a 
cylindrical volume coated on the inside with reflective paint (95\% reflectivity) that 
significantly increases the detector's light collection. 
We demonstrate several calibration techniques using point sources 
and cosmic-ray muons for energies up to 20 MeV for this large liquid scintillation detector. Neutron-gamma separation 
using pulse shape discrimination with a few MeV neutrons to 
hundreds of MeV neutrons is shown for the first time using a large liquid scintillator. 
\end{abstract}

\begin{keyword}
liquid scintillator \sep neutron detection \sep underground experiments
\PACS 25.30.Mr \sep 28.20-v \sep 29.25.Dz \sep 29.40.Mc
\end{keyword}
\end{frontmatter}
\section{Introduction}
The Sanford Underground Research Facility (SURF) was chosen as a 
site for ultra-low background experiments. The current two experiments are the direct detection of 
dark matter utilizing xenon with the Large Underground Xenon (LUX)~\cite{lux} experiment and the search for 
neutrinoless double-beta using germanium with the M{\sc ajorana} D{\sc emonstrator}~\cite{majorana}. 
Understanding the in-situ background levels is extremely important for these rare-event physics
experiments. Although positioning experiments in a deep underground laboratory significantly 
suppresses the background caused by cosmic-ray muons, the residual muons still create fast 
neutrons~\cite{meihime}. 
The intensity of the muon-induced neutrons depends largely on the depth of the underground laboratory~\cite{meihime}.
The energy spectrum, multiplicity, and angular distribution of the muon-induced neutrons are not particularly well measured. 
In addition, there are also fast neutrons from ($\alpha$,n) reactions that are 
created in the surrounding rock by natural radioactivity.
In this paper we demonstrate a neutron background characterization technique that 
has been developed for a large liquid scintillation detector. 
\par
One major problem in neutron detection 
is the separation of neutrons from the electromagnetic background caused by gamma rays from the environment and  internal 
contamination of the detector materials. The pulse shape discrimination technique, which uses the difference in the shape of the 
scintillation pulses 
generated by neutrons and gamma rays, has been implemented
successfully with small neutron detectors for many 
years~\cite{icarus, bose, normand, soderstrom, bell}. 
Unfortunately, small neutron detectors (of a few liters or less) have low efficiency when detecting neutrons and are
 high cost compared to size and efficiency. 
For the successful neutron background 
characterization, we require a higher efficiency of neutron detection
and a broader energy sensitivity, along with lower cost per detector. The development of a relatively large volume 
detector increases the possibility of significantly improving the neutron detection efficiency. 
More importantly, it opens a window for exploring the neutron
energy at a few MeV up to a few hundred MeV.  
 Although most of the neutron-gamma discrimination experiments 
were carried out using small detectors,
 the possibility of neutron-gamma discrimination using time of flight (TOF) measurements
have also been investigated using a large neutron detector~\cite{ito}. 
In this work we investigate the possibility of using a large volume detector for the direct 
detection of neutrons with energy of a few ten MeV, with pulse shape discrimination to distinguish from gamma rays.
\section{Experimental setup}
\begin{figure}
\includegraphics[angle=0,width=\textwidth]{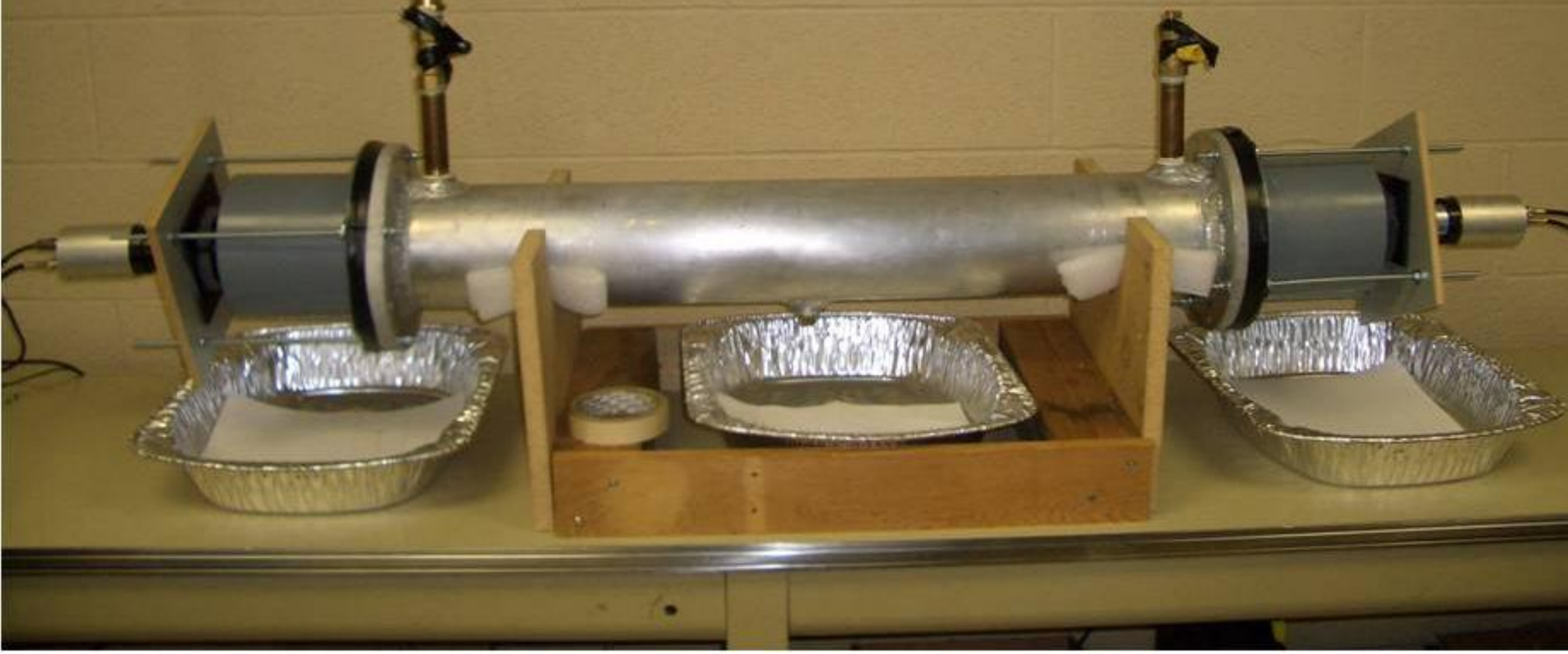}
\caption{\label{geometry} The setup of our liquid scintillation detector. The aluminum foil pans are required for safety in case of leaks. }
\end{figure}

 We have constructed a  
liquid scintillation detector that is 
1 meter long and 5 inches (12.7 cm) in diameter. 
This detector is fabricated using a aluminum cylindrical housing 
with two Pyrex windows on each side of the cylinder attached to PMTs. 
To mitigate light loss, the inner surface of the detector was 
covered with reflective paint EJ520 (Eljen Technology) that has 95\% reflectivity. The reflective paint 
makes scintillation photons scatter multiple times from the detector walls, which 
partially compensates for the relatively poor PMT photo-cathode coverage. The detector volume was filled 
with liquid scintillator EJ301, which is  specifically designed for neutron-gamma discrimination. 
Scintillation light is collected by two 5-inch Hamamatsu R4144 PMTs attached to both 
Pyrex windows. Optical grease is used to couple the PMTs with the Pyrex
windows to avoid optical mismatch and reflections. 
The geometry of the detector setup is presented in Fig. \ref{geometry}. 
The total photo-cathode coverage of the detector is at about 3\% and the PMT 
quantum efficiency is about 20\% for photons with a wavelength of 300-700 nm. The EJ301 is composed of 
carbon and hydrogen atoms, and has a H/C ratio of 1.212 and a density of 0.874 g/cm$^{3}$. 
Its light output is 78\% as of anthracene with a maximum 
emission at 425 nm, which exactly corresponds to the most sensitive region of the Hamamatsu PMTs. 
\par
\begin{figure}
\includegraphics[angle=0,width=\textwidth]{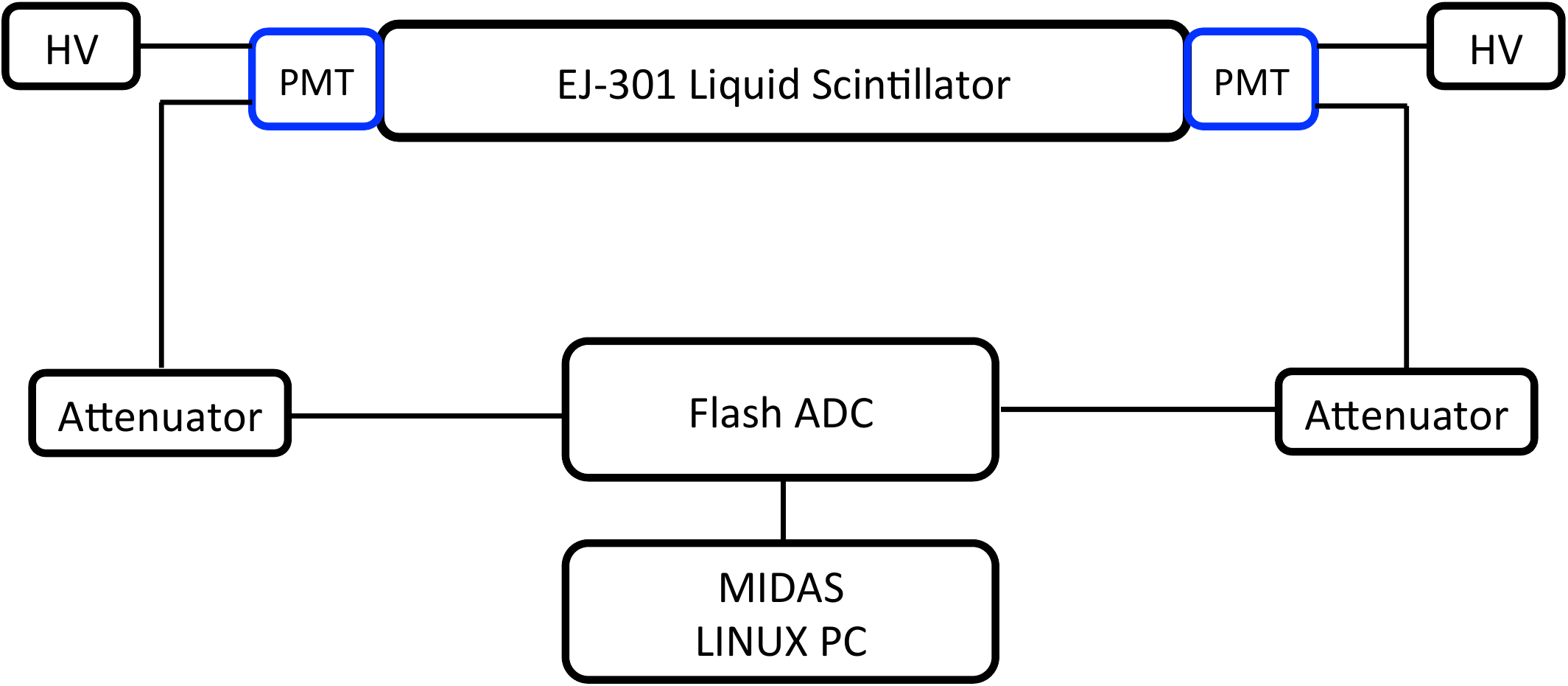}
\caption{\label{electronics} The electronic system of the detector.}
\end{figure}

Contamination of oxygen in the liquid scintillator results in a reduction of light output. 
In order to minimize the amount of oxygen contamination, 
the scintillator was thoroughly purged with dry argon 
and then the detector volume was sealed before the measurements were performed. 
The operational voltages of both PMTs were determined to be 2000 V at which the PMT gain is 1.4$\times$10$^{6}$\cite{hamamatsu}.   
The DAQ consists 
of a fast flash ADC that analyzes the PMT output signal. 
The sampling frequency of the flash ADC 
is 170 MHz, providing a signal amplitude every 5.88 ns.
The maximum amplitude the ADC can handle is 4096. If the peak sample of an event extend to be greater than
4096, we categorize it as a saturated event. 
In order to avoid saturation from high
energy events, especially those close to the PMTs, both PMT output signals are attenuated at
23 dB before connecting to the DAQ.
The pedestal level of the ADC is found to be $\sim 1295$ for PMT0 and $\sim 1267$ for PMT1 under the high
voltage of 2000 V.
The ADC is controlled by a
program based on the MIDAS data acquisition system software~\cite{midas}.
The electronic system is shown in Fig. \ref{electronics}.
\section{Energy and position calibration procedures}
\begin{figure}
\includegraphics[angle=0,width=\textwidth]{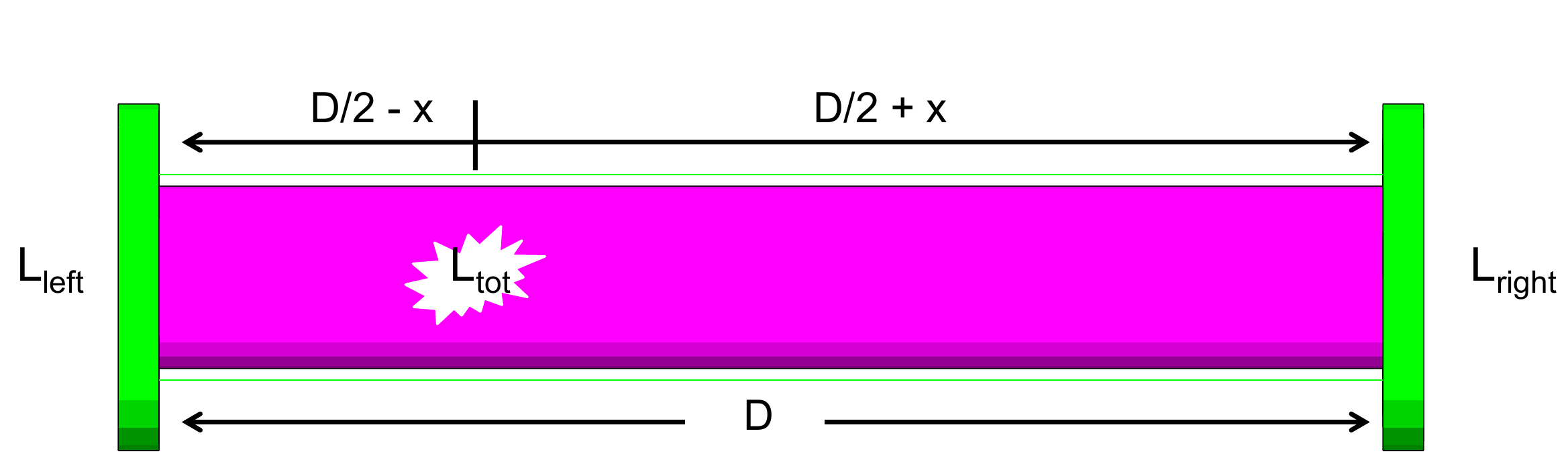}
\caption{\label{independentDiagram} Position of the incident
                particle and the total light yield being split
                and collected by individual PMTs.}
\end{figure}

For a large scintillator, the detector response to energy with each individual 
PMT is determined by the energy 
deposition and the position where the incident particles interact with a target in the scintillator.
This position dependence is caused by the attenuation of the light 
in the tube~\cite{kims}. 
To obtain the position information, the charge ratio from
the two PMTs is used to characterize the position of an incident particle. The following criteria are applied to select events:
\begin{enumerate}
\item Both PMTs must be triggered. 
\item Both signals must not saturate the ADC.
\item Time coincidence is within 30 ns, which is the time difference of the 
	largest sample in the pulse between two PMTs.  
\end{enumerate}
Assume an energy deposition of $E_{tot}$ is created at a distance of $X$ (distance to the middle,
see Fig. \ref{independentDiagram}.), the light collection by two PMTs $L_{left}$
and $L_{right}$, and the total light yield $L_{tot}$ 
proportional to $E_{tot}$. Without any energy loss, the total light yield
would be evenly split by the two PMTs,
$L_{left} = L_{right} = 0.5L_{tot}$. In reality we have to consider
light attenuation, especially for such a big detector.
Taking $l$ as the attenuation
length in the scintillator and $D$ as the total length of the tube,
a simple calculation in Eq.(\ref{eq:position}) shows that
the position can be determined by the combination of $L_{left}$ and $ L_{right}$, respectively.
\begin{eqnarray}\label{eq:position}
L_{left} &=& 0.5L_{tot}e^{-(D/2-X)/l},  \nonumber \\
L_{right} &=& 0.5L_{tot}e^{-(D/2+X)/l},  \nonumber \\
\ln{\sqrt{\frac{L_{left}}{L_{right}}}} &=& X/l.
\end{eqnarray}
The light collection $L_{left}$ and $L_{right}$ will then be converted to photoelectrons at the photocathode of the PMTs. 
If we define $a0$ to stand for the total charge converted from $L_{left}$ and $a1$ the total
charge converted from $L_{right}$, $a0$ should be proportional to 
$L_{left}$ ($a0\propto L_{left}$) and $a1$ proportional to $L_{right}$ ($a1\propto L_{right}$).
Thus, we can use the charge ratio, $\ln{\sqrt{a0/a1}} \propto X/l$, 
 to interpret the position of the particle along the tube. 
\par
\begin{figure}
\includegraphics[angle=0,width=\textwidth]{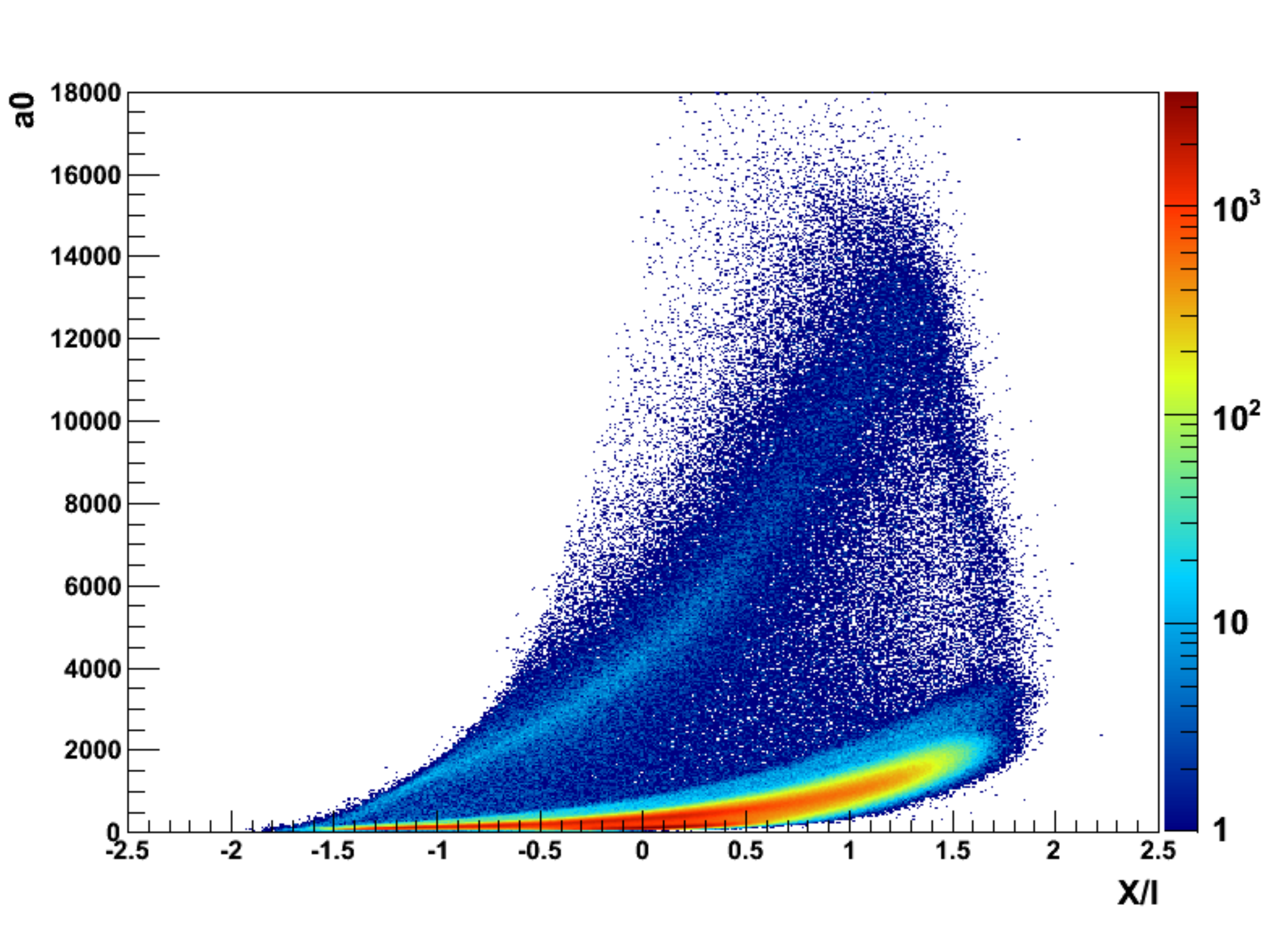}
\caption{\label{detResponse} Detector response to energy with respect to one of the PMTs for a surface background run.
		$X/l$ stands for the relative position of the particle. $a0$ is the 
		integrated charge (the pulse area) of the PMT0.    
		}
\end{figure} 
The detector response to energy with respect to one of the PMTs for a surface 
background run is presented in Fig. \ref{detResponse}.
The value of $X/l$ varies from -2 to 2 which is four times the attenuation length 
from one end of the tube to the other.    
Considering the length of the tube is 1 meter, a mean attenuation
length of $\sim$25 cm is determined for the light transport in the scintillator. 
Note that this result is much less than the expected attenuation length for a
EJ301 scintillator ($l>1$ meter). This is because the the light transport in the tube is 
dominated by the diffusive reflection, which dramatically exacerbates the
attenuation.  
\par 
In the lower energy range, the detected events are dominated by gamma rays from 
the internal contamination of the detector components. 
In contrast, events at the high energy range are mainly
from the cosmic muons. Clearly, the higher energy curve in 
Fig. \ref{detResponse} indicates a muon minimum 
ionization peak of $\sim$20 MeV at a 5-inch diameter. This minimum ionization has also been 
verified using a GEANT4 simulation~\cite{geant4}.
The incident cosmic ray muons should be uniformly 
distributed along the tube (except for 
edge effects at both ends of the tube), which 
provides us a natural energy calibration source for energy up to 20 MeV. 
This curve shows that the energy response (total charge collected by PMT) of a single PMT 
has strong position dependence.
Within the tube, i.e., $X/l\in[-2,2]$, there are three blocks filled with no data. 
The blank area at the bottom in Fig. \ref{detResponse} is caused 
by the energy threshold set for the trigger of the PMTs while the blank areas on both sides 
are caused by the saturation of the flash ADC. 
\par  
Energy-scale calibration is performed using a $^{22}$Na gamma ray source. 
The $^{22}$Na radiation source 
produces two gamma ray lines with 0.511 MeV and 1.275 MeV energies. Measurement is performed by setting 
the uncollimated source on the top of the aluminum tube every 2.5 cm 
from one end to the other.  
Calibration is made 
using the 1.275 MeV gamma ray line from a $^{22}$Na source.  
\begin{figure}
\includegraphics[angle=0,width=\textwidth]{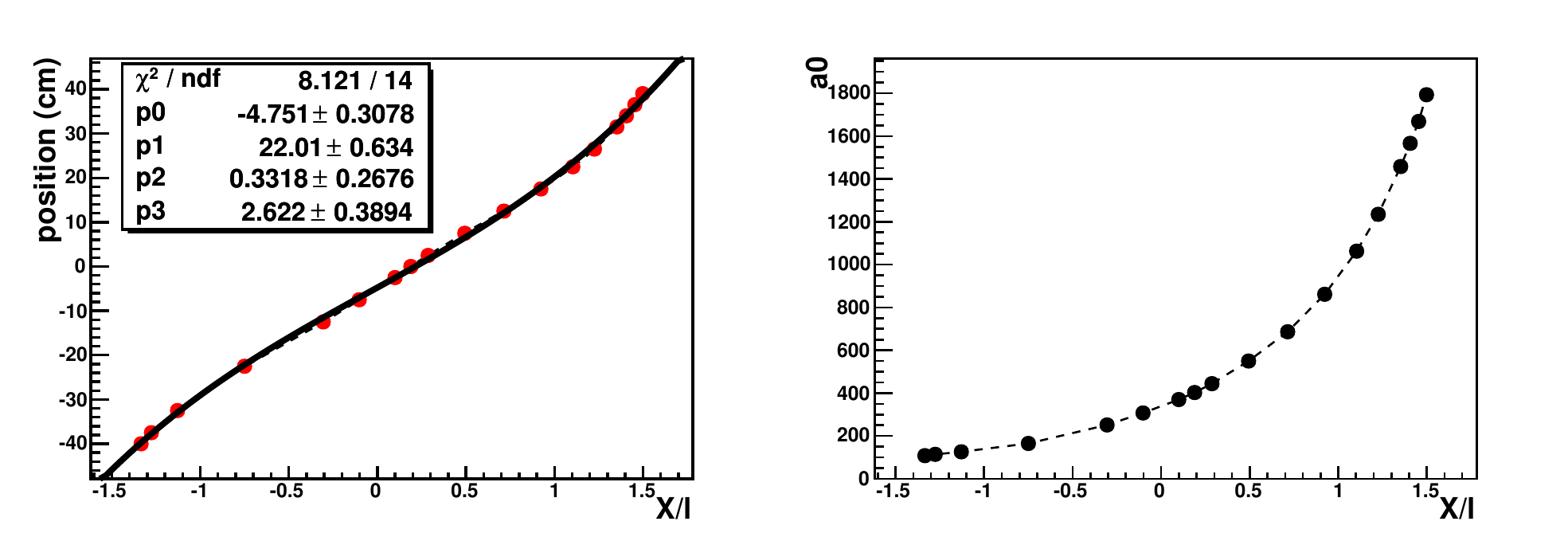}
\caption{\label{responseToNa22} The position (left) and energy (right) calibration 
			from a $^{22}$Na source in terms of the
			1.275 MeV gamma ray line. The data in the left plot is
			fitted by a third-order polynomial function. }
\end{figure}
The left plot of Fig. \ref{responseToNa22} gives a position calibration interpreted by
$X/l$. It results in nearly linear relation between $X/l$ and
the actual position. Note that the curve doesn't pass through the central point(0, 0) 
because the gains of two PMTs are not tuned to be identical which caused the asymmetry of 
the detector response.  
The right plot shows 
a position-dependent energy response to the  1.275 MeV 
gamma rays from the source. 
\par
\begin{figure}
\includegraphics[angle=0,width=\textwidth]{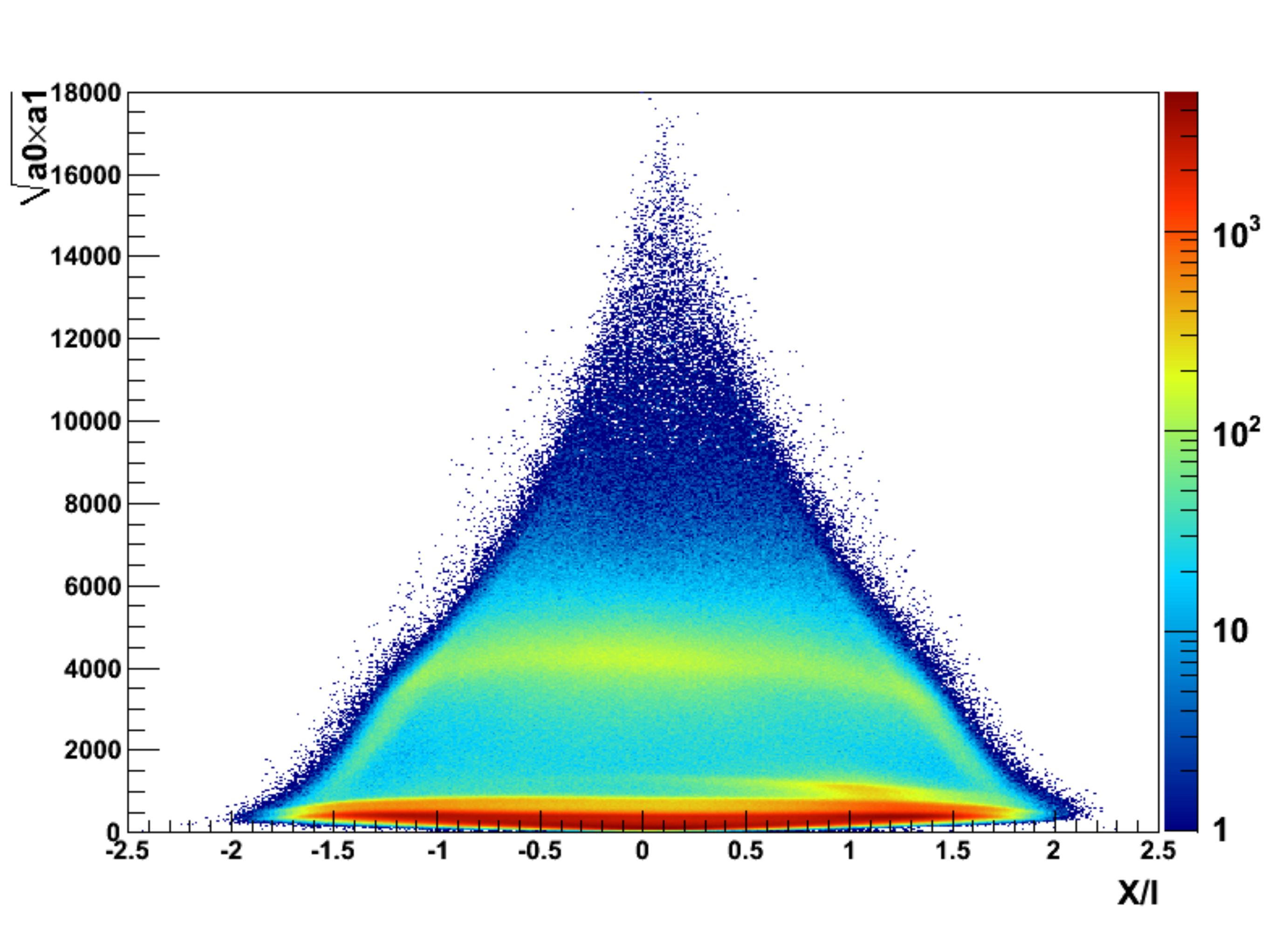}
\caption{\label{posiIndependent} Plot of $\sqrt{a0\times a1}$ vs $X/l$ to
		show the position independence of energy along the tube. 
		This is a surface run where an AmBe source was placed on top of  
		the tube and 22.5 cm from the right end.
		}
\end{figure}

The position dependence of energy makes the energy calibration complicated. 
Since both ends of the aluminum tube are coupled to PMTs, 
the energy scale can be
position independent~\cite{LGB} based on the Eq.(\ref{eq:independence}). 
\begin{equation}\label{eq:independence}
\sqrt{L_{left}L_{right}} = 0.5L_{tot}e^{-D/2l} \propto E_{tot}.
\end{equation}
Since $\sqrt{L_{left}L_{right}}$ is proportional to $\sqrt{a0\times a1}$, we can use $\sqrt{a0\times a1}$ to 
represent the energy scale and 
avoid the position dependence along the tube. To verify this hypothesis, 
a plot of $\sqrt{a0\times a1}$ versus $X/l$ has been created
in Fig. \ref{posiIndependent} where  a distinct and  
almost horizontal line is shown for the muon minimum ionization process 
with the $\sqrt{a0\times a1}$ at $\sim$ 4000. 
Other than the edge effect at both sides, it shows that
the position independence of energy is quite accurate if we use $\sqrt{a0\times a1}$
to describe the total energy deposition.     
Using $^{22}$Na and AmBe~\cite{geiger} sources we also obtain consistent results 
for gamma ray lines at 1.275 MeV and 4.4 MeV.
Therefore, an energy calibration 
curve is obtained for few MeV up to 20 MeV.    
\section{Neutron-gamma discrimination}
\begin{figure}
\includegraphics[angle=0,width=\textwidth]{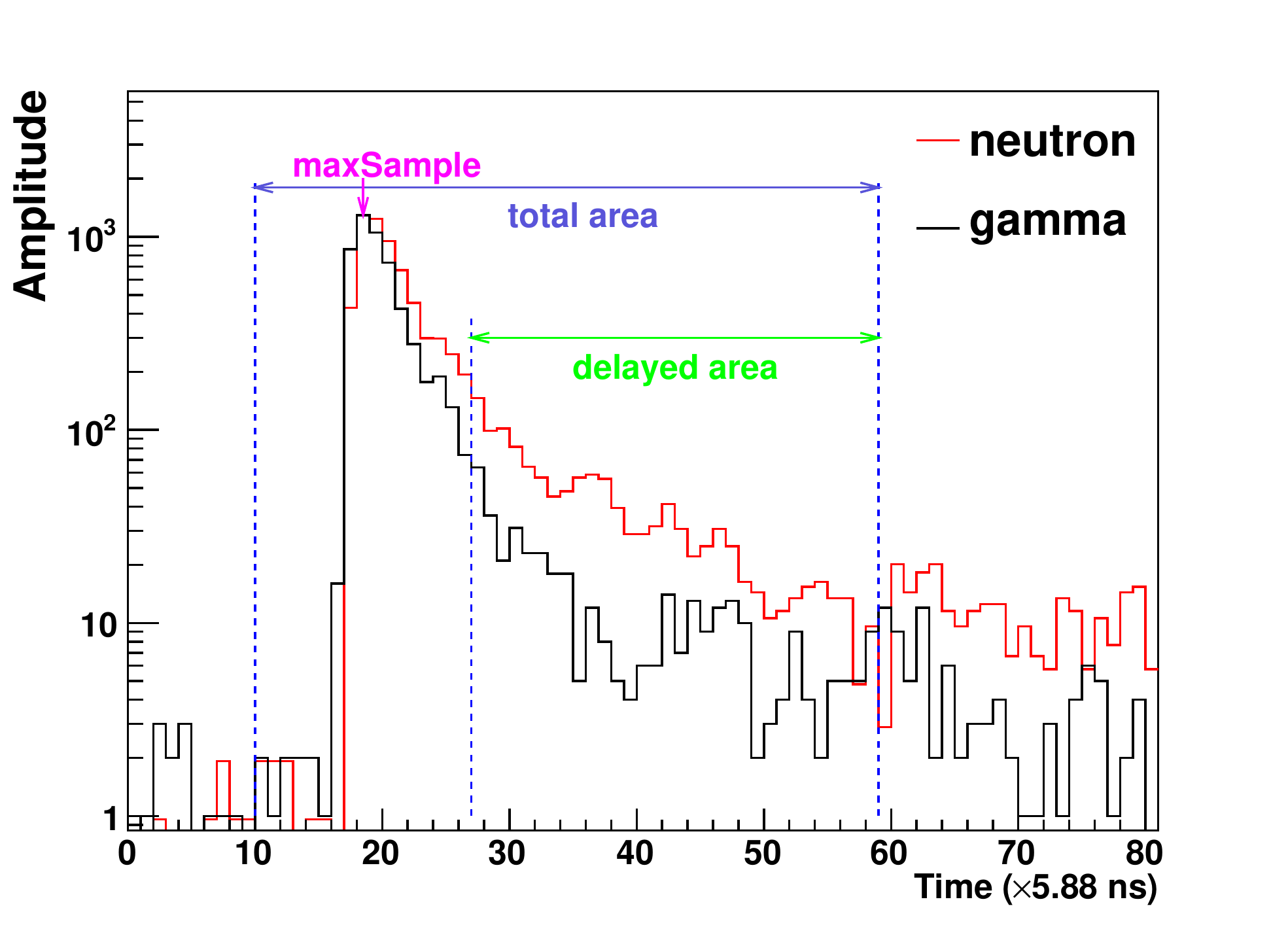}
\caption{\label{areaDefine} A comparison of the pulse shape between neutron and gamma events. The pedestal 
with the amplitude of 1295 is subtracted from each pulse sample. }
\end{figure}

Measurements of neutron-gamma discrimination  were 
carried out using an AmBe neutron source~\cite{geiger}. 
It was placed vertically 6.5 cm directly above the outer wall of 
the tube and horizontally 22.5 cm away from PMT0. The neutron 
emission rate of this source is $\sim$100 Hz, and the neutron energy 
spectrum is up to 11.2 MeV. 
The digitized pulses from each event have been recorded for 
data analysis. The analysis procedure includes calculating the total charge per pulse per event  and the charge 
that corresponds to the tail of the pulse (delayed charge). 
The total charge is defined as the integral under the pulse
from 8 samples before to 40 samples after the peak, 
where the pulse sample is defined as the signal amplitude every 5.88 ns. 
The delayed charge is optimized to be the integral under the pulse from
8 to 40 samples after the peak(see Fig. \ref{areaDefine}).
Since the decay-time constants for scintillation light from neutrons and gamma rays are different, 
the neutron-gamma discrimination can be obtained
using the ratio of the delayed charge to the total charge, 
$\frac{d(A)}{t(A)}$, where $d(A)$ is the delayed charge integrated over the delayed pulse area and $t(A)$ 
is the total charge integrated
over the total pulse area. 
\par
\begin{figure}
\includegraphics[angle=0,width=\textwidth]{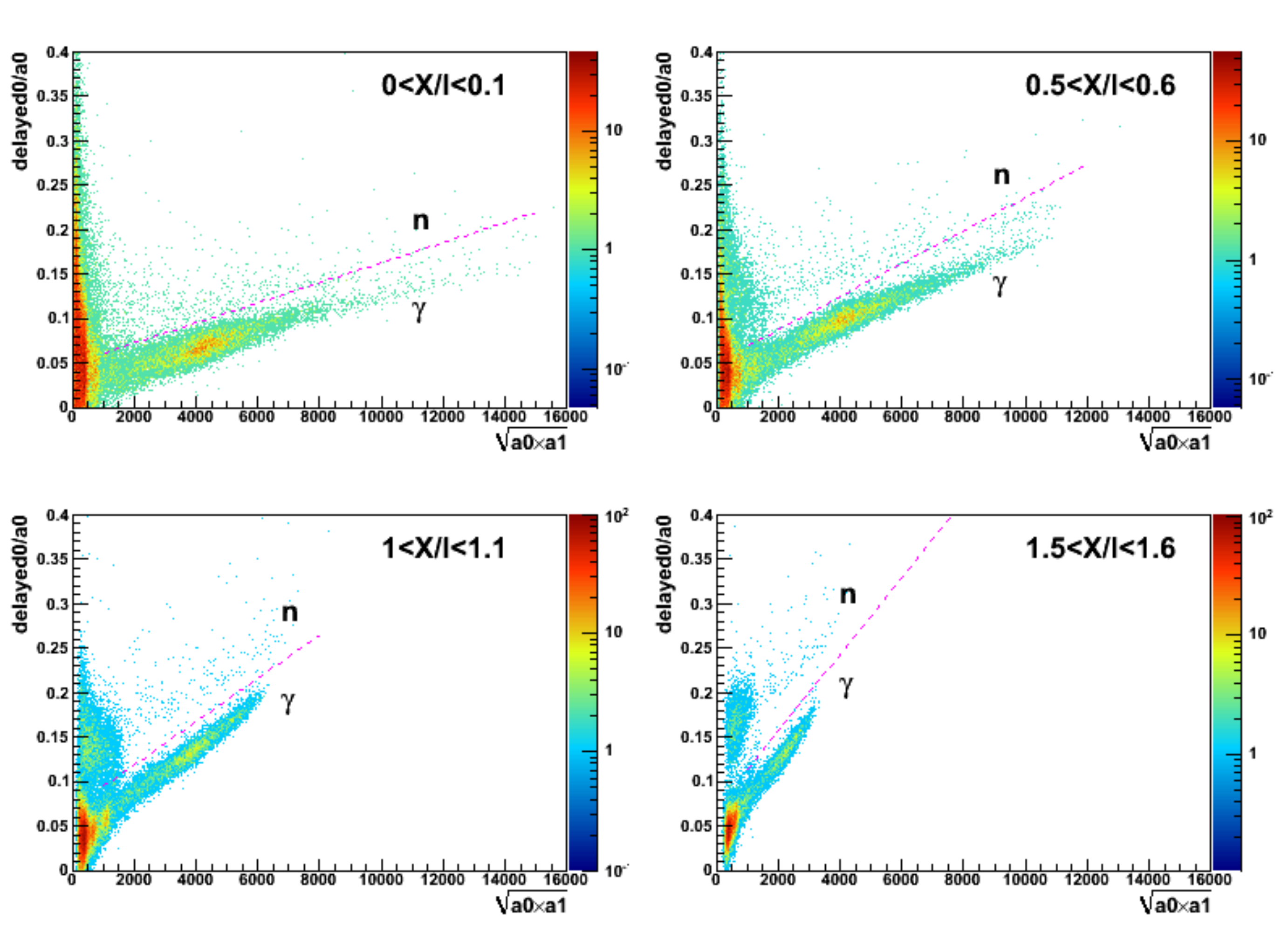}
\caption{\label{ngSlice} The performance of neutron-gamma discrimination from PMT0. 
			 Relative position $X/l$ at 0, 0.5, 1 and 
			1.5 are displayed in the individual plots.}
\end{figure}
\begin{figure}
\includegraphics[angle=0,width=\textwidth]{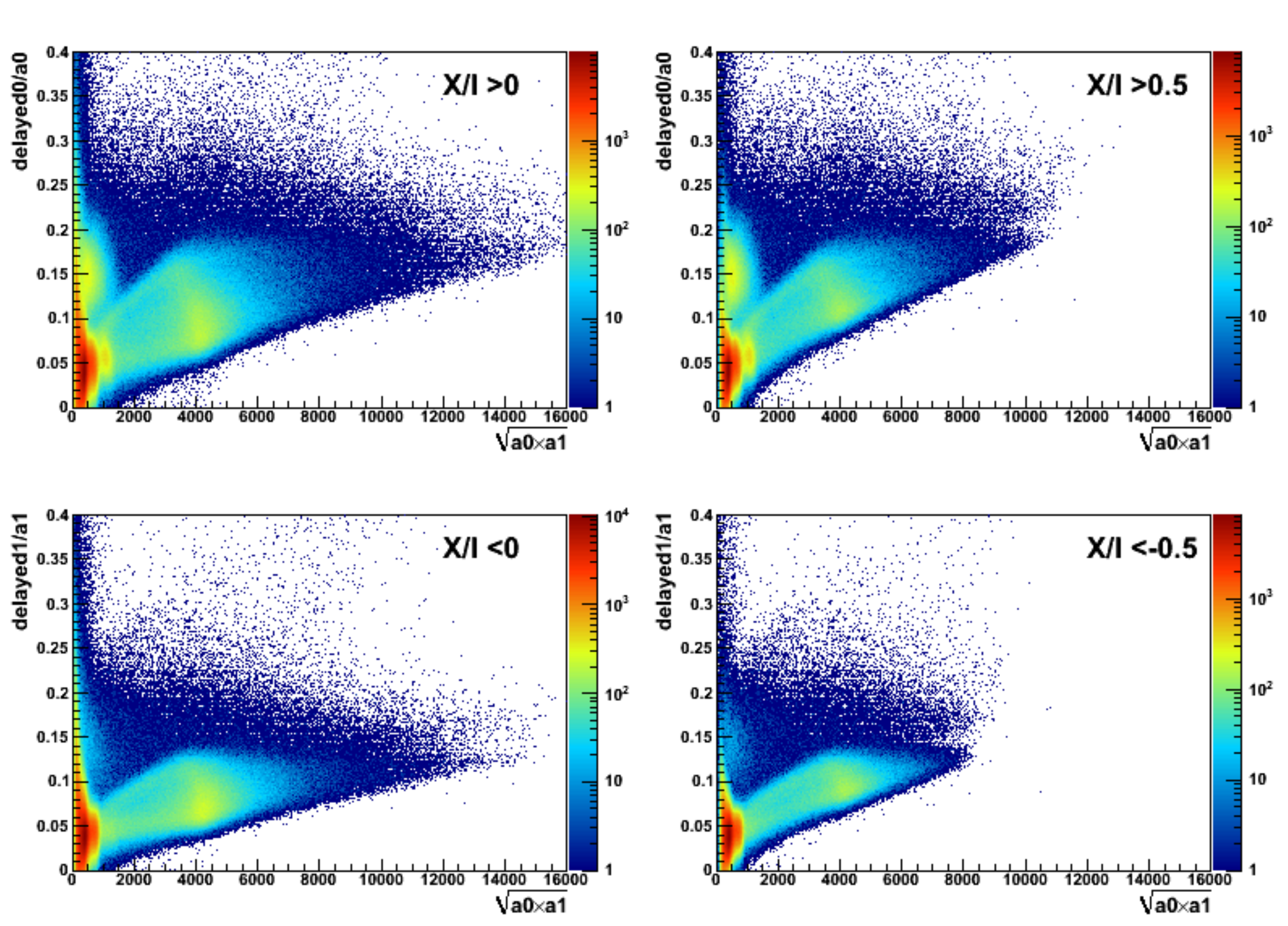}
\caption{\label{ngBuried} Neutron-gamma discrimination for a broad
		position range. The top two plots adopt the signals from PMT0
		at position ranges $X/l>0$ and $X/l>0.5$ while the bottom
		two plots take the signals from PMT1 at position ranges
		$X/l<0$ and $X/l<-0.5$.  
		}
\end{figure}

A dedicated study shows that the neutron-gamma separation 
can only be found in a very narrow range of position along the tube (see Fig.  \ref{ngSlice}).
The plots describe the separation of neutrons and gamma rays in terms of their 
pulse shape differences. 
It clearly shows that neutron-gamma separation decreases when
energy deposition occurs far from the target PMT. This is expected
because the diffusive reflection, absorption and re-emission of light 
washes out the pulse shape difference between neutron and gamma ray events, 
which happen farther away from the target PMT. 
Therefore the charge ratio $delayed1/a1$ from PMT1 
is adopted to separate the neutrons at one end of the tube ($X/l<0$) and 
$delayed0/a0$ from PMT0 is used for the other end ($X/l>0$).   
The neutron-gamma separation vanishes for a broad position range as shown
in Fig. \ref{ngBuried}.

\begin{figure}
\includegraphics[angle=0,width=\textwidth]{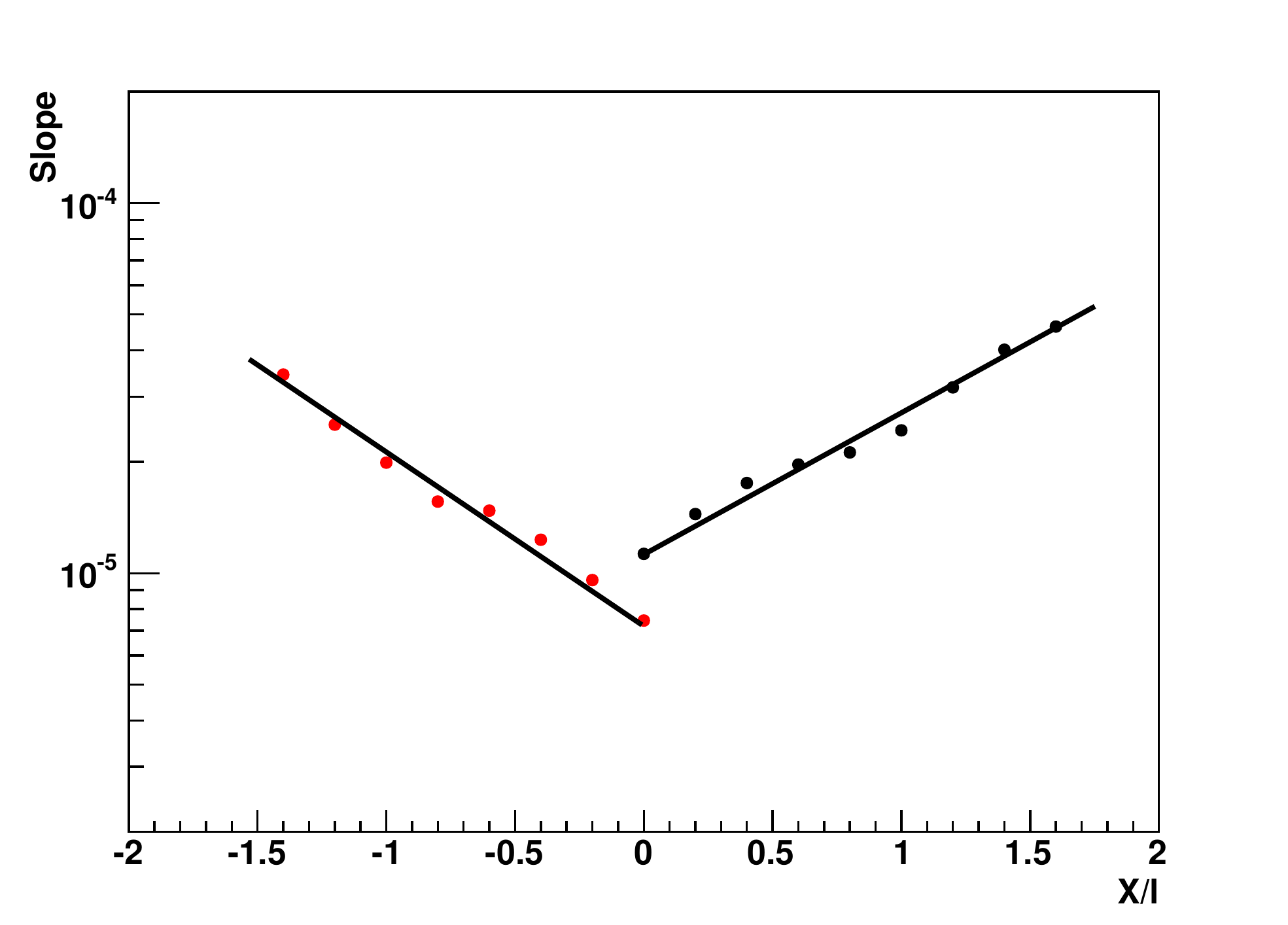}
\caption{\label{posi_slope} The relation between the position and slope of 
		the gamma ray band. 
                }
\end{figure}

The plots in Fig. \ref{ngSlice} indicate that the gamma ray band and the neutron band
are separated when position is held constant. However, they have different slopes
as the position varies, which results in the separation disappearing when we use 
a wide position range. 
Since $\sqrt{a0\times a1}$ represents energy, which is found to be independent of
the position, the energies ($x$-axis) of the invariant data points are kept and
the value of $y$, or ratios of the charge between the delayed and the total are projected
 to the horizontal line for all data points (according to the slope $k$, i.e.,
$(x, y)\Rightarrow(x, y-kx)$). The relation between the position and the slope 
of the gamma ray band is interpreted in Fig. \ref{posi_slope}. 

\begin{figure}
\includegraphics[angle=0,width=\textwidth]{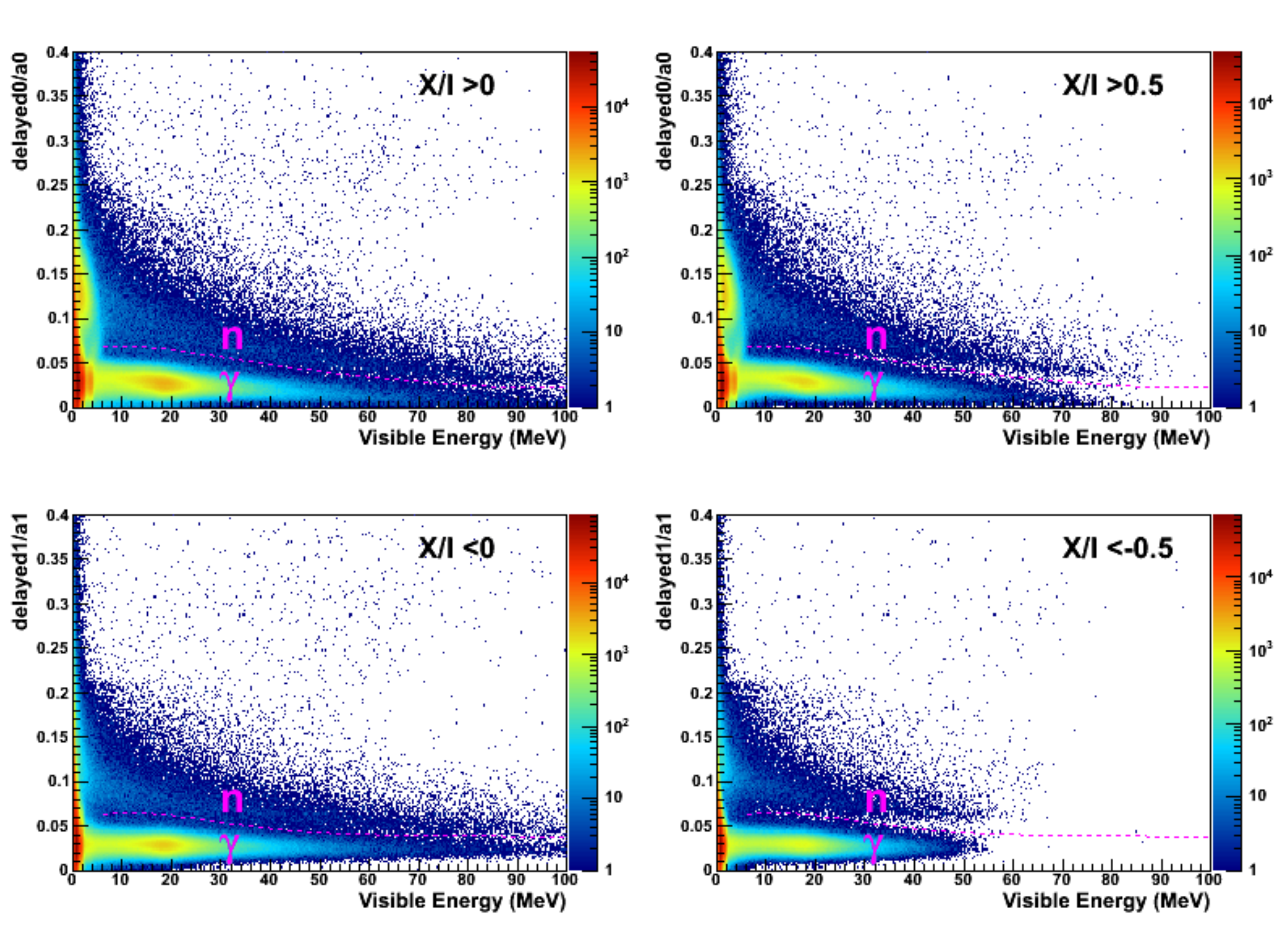}
\caption{\label{ngCorrected} The corrected neutron-gamma separation results in terms of 
		signals from PMT0 (top two) and  PMT1 (bottom two). 
		The energies are also corrected based on calibration.
		}
\end{figure}

\begin{figure}
\includegraphics[angle=0,width=0.499\textwidth]{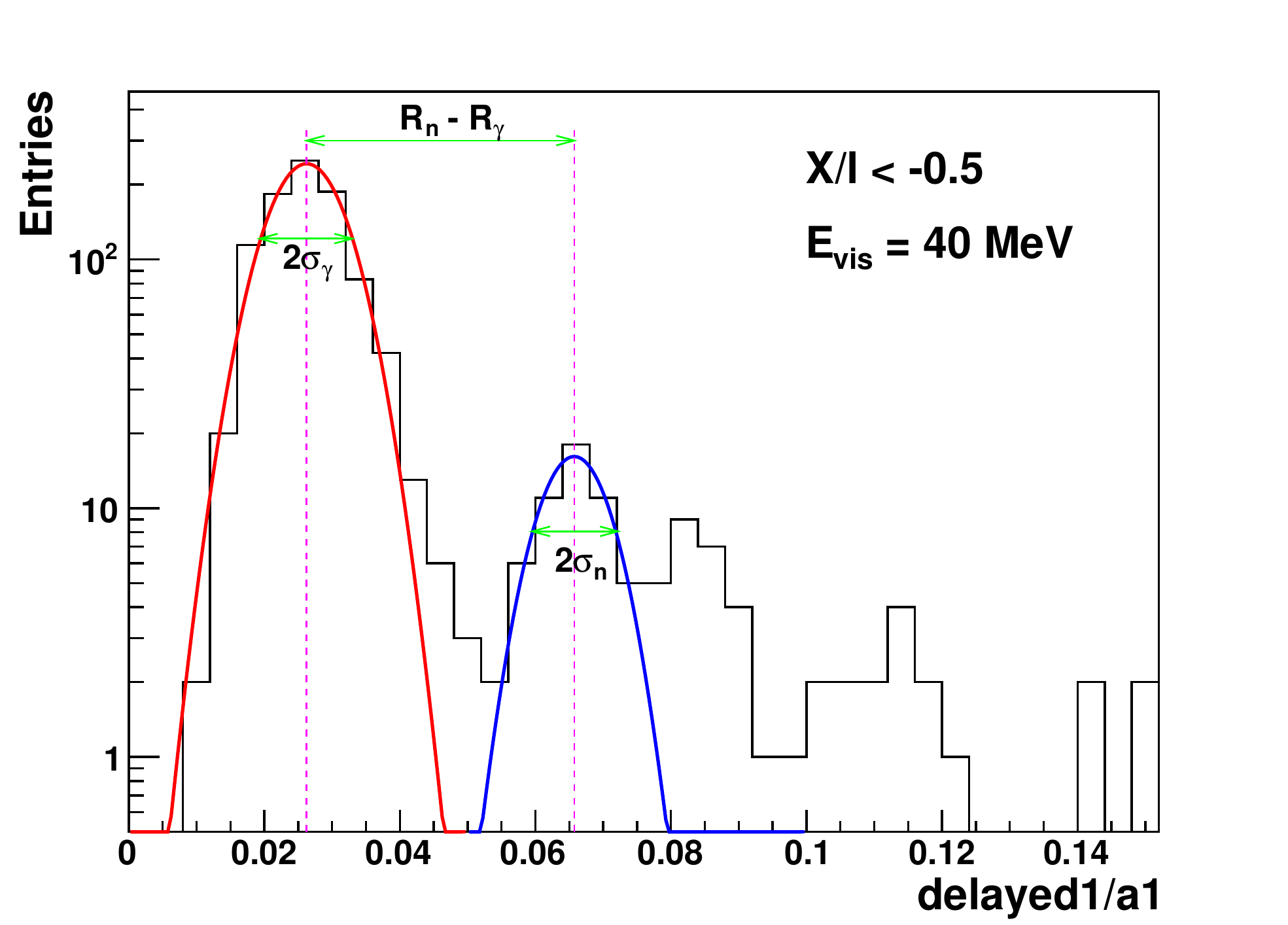}
\includegraphics[angle=0,width=0.499\textwidth]{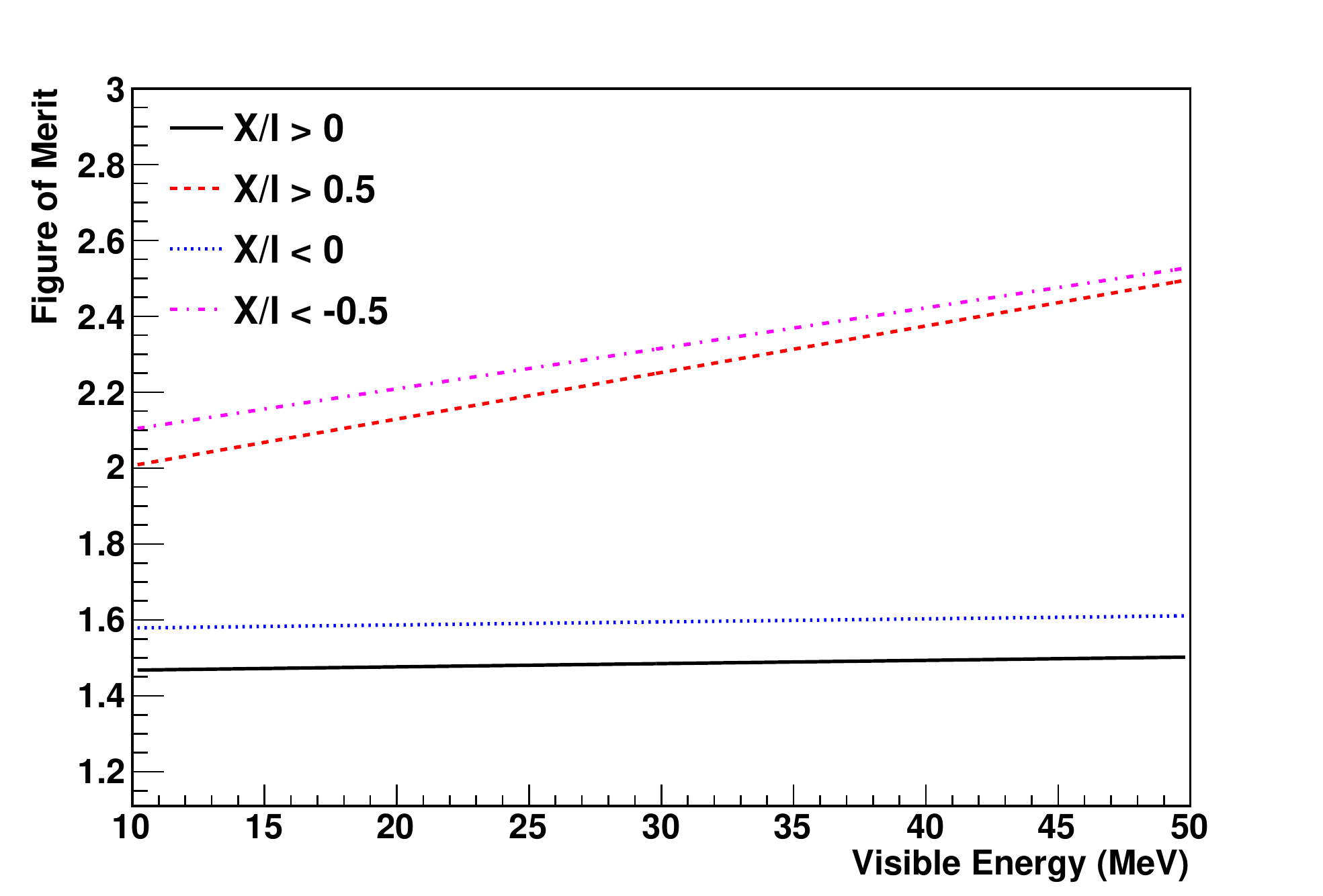}
\caption{\label{merit} The left plot is a projection on the charge ratio, 
		$delayed1/a1$, for the position range $X/l<-0.5$ and the energy
		at 40 MeV. The right plot shows the comparison of the 
		Figure of Merit that characterizes the goodness of 
		separation between neutrons and gamma rays versus the visible
		energy.  
                }
\end{figure}

After summing all the 
narrow position slices together, a combined neutron-gamma separation plot is obtained
as shown in Fig. \ref{ngCorrected}. In order to get better neutron-gamma separation, the condition 
$|X/l|>0.5$ is required to limit the positions to values that are not  too far away from the PMTs.   
In order to compare the improvement of the separation, the figure of merit~\cite{fom} that describes
the goodness of separation between gamma rays and neutrons is defined below:
\begin{equation}\label{eq:merit}
	Goodness=\frac{R_{n} - R_{\gamma}}{\sigma_{n}+\sigma_{\gamma}},
\end{equation}
where $R$ is a mean of the distribution for the ratio between the delayed charge and the total
charge and $\sigma$ is the Half Width at Half Maximum (HWHM) of the distribution of the charge ratio.
The left plot in Fig. \ref{merit} illustrates how the goodness of separation is calculated for 
the position range $X/l<-0.5$ and the energy at 40 MeV. A Gaussian distribution is assumed 
for both neutrons and gamma rays. The multiple peaks in the neutron band indicate different 
recoils induced by protons, alphas, and $^{12}$C. 
The right plot in Fig. \ref{merit} gives  
the comparison of the figure of merit before and after the 
position cut, which is illustrated in Fig. \ref{ngCorrected}. 
It shows that the neutron-gamma separation gets better after position cuts are applied. 
\par
\begin{figure}
\includegraphics[angle=0,width=\textwidth]{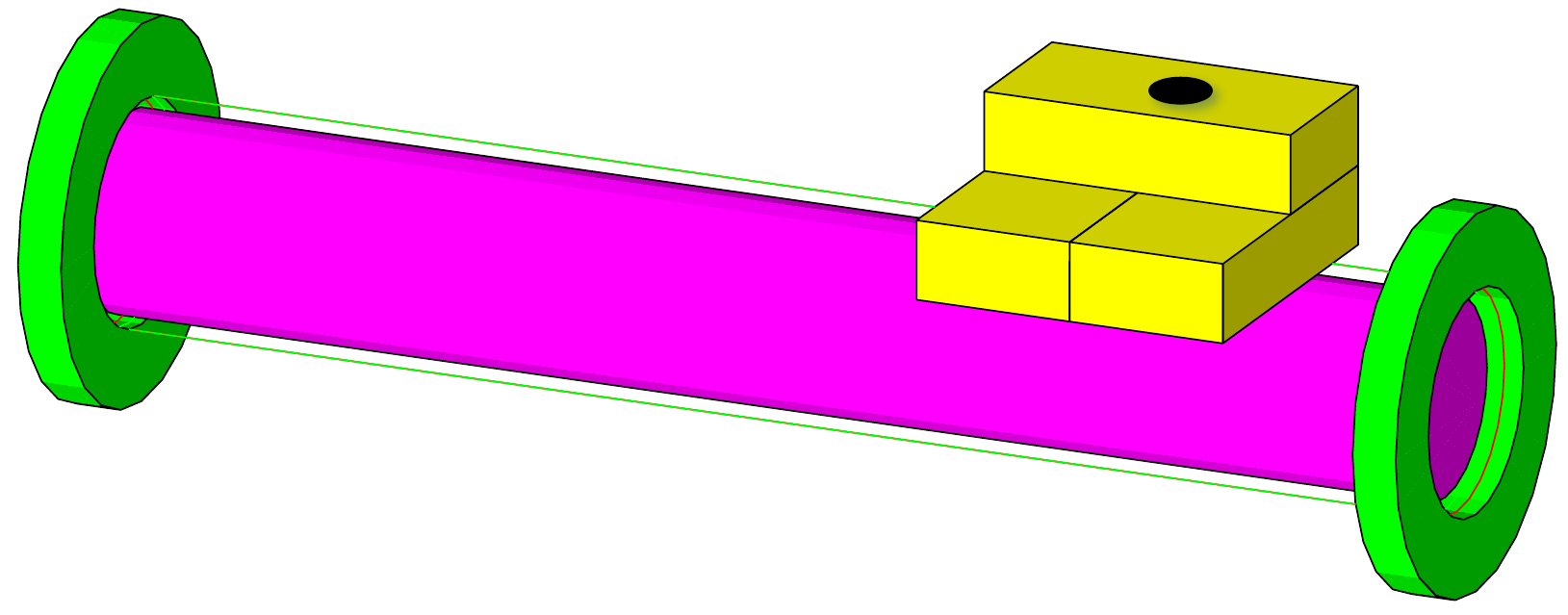}
\caption{\label{Setup_AmBeWithLead} Setup for the run with an AmBe source at 
			Soudan Mine underground. The source is located 22.5 cm 
			away from the right hand PMT and above two layers
			of lead bricks (total 4 inch). 
			}
\end{figure}

 
Given the neutron energy spectrum of the AmBe source, the maximum neutron recoil energy detected by 
the detector is up to 6 MeV. However, for a surface run,
the cosmogenic neutrons can create higher energy events.
Because the AmBe source is much closer to PMT0 than PMT1, we 
see more source related events in the upper two plots of Fig. \ref{ngCorrected}. 
There are distinct peaks in the gamma-ray band, which are the muon minimum 
ionization peaks for $\sim$ 20 MeV and 4.4 MeV gamma-ray lines from AmBe source. 
Although the muon ionization peak looks diffused in a wide range due to the 
edge effect, it gives a well defined peak once we remove the range that is close
to the PMT ends. The neutron-gamma separation performs well up to 
100 MeV, which indicates this detector is able to 
detect neutrons with energies of up to a hundred MeV.   
The low energy range ($<6$ MeV) 
of its neutron band is overwhelmed by AmBe neutrons. However, this cannot be 
separated from the gamma-ray band due to the contamination of 4.4 MeV gamma rays from the AmBe source.
This is because some gamma rays associated with neutrons enter into the detector at the same
time. In order to remove this source related gamma-ray contamination, 4 inch lead
bricks were added between the AmBe source and the detector. The whole setup is simulated using GEANT4
and its geometry is displayed in Fig. \ref{Setup_AmBeWithLead}.
\par
\begin{figure}
\includegraphics[angle=0,width=\textwidth]{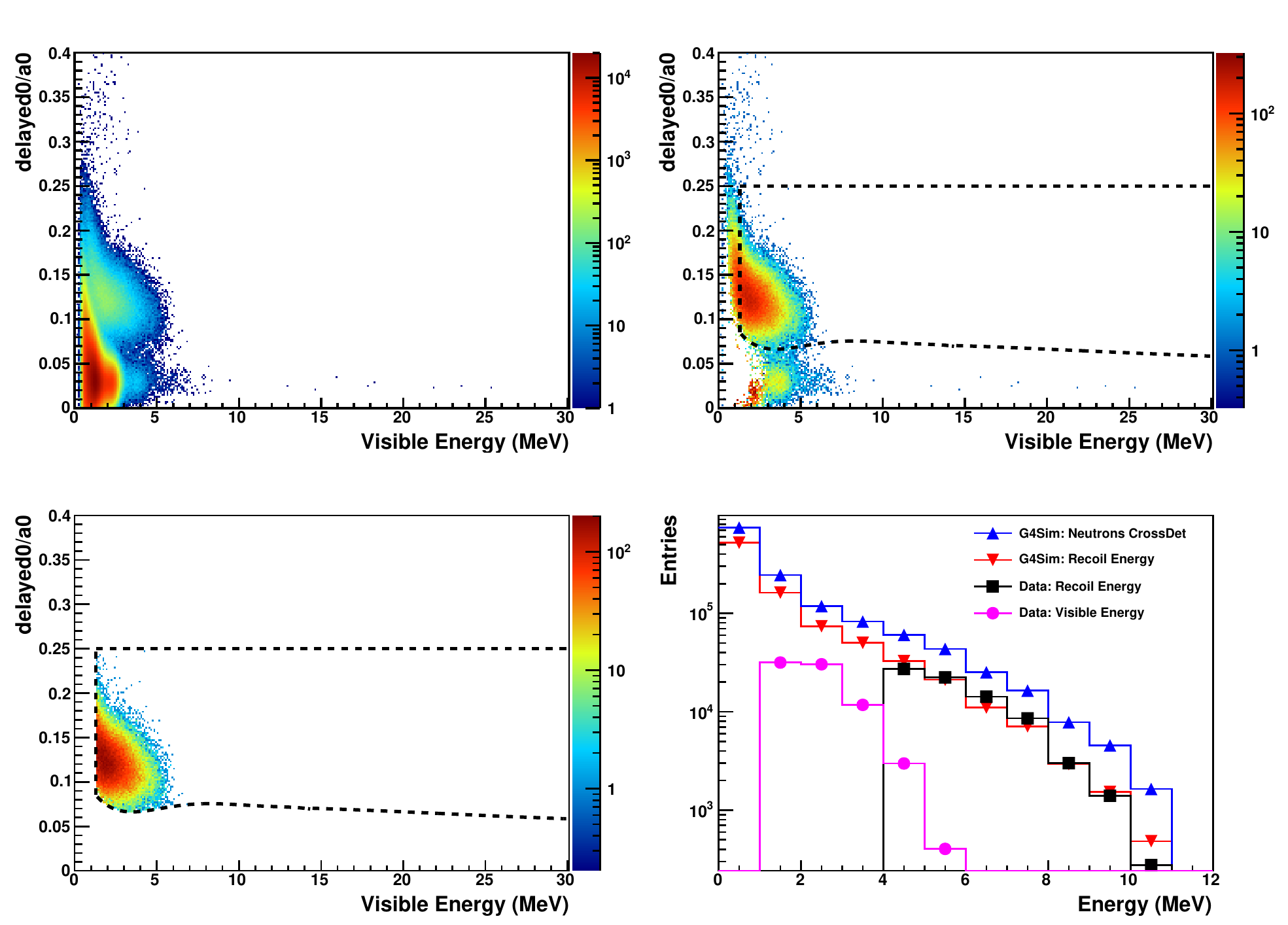}
\caption{\label{ambeunderground} A calibration with an AmBe source at the Soudan mine underground.
		The upper left plot is a combination of the AmBe source signal and
		background data. The upper right plot gives source signal with
		background subtracted. The bottom left plot shows neutrons
		separated from gammas at 1.3 MeV energy and for a
		$delayed0/a0<0.25$ charge ratio cut. The bottom right plot
		displays the processes for neutron energy reconstruction.
		The result is compared with simulation.   
 		}
\end{figure}

Rather than implementation on the surface, the new AmBe run 
was performed underground at Soudan mine (2100 m.w.e).
This has the extra benefit of dramatically reducing the high
energy backgrounds from cosmic rays. 
The entire process of data analysis and reconstruction is shown in 
Fig.  \ref{ambeunderground}.
Comparing the upper right plot in Fig. \ref{ngCorrected} to the upper left plot
of Fig. \ref{ambeunderground}, one can clearly see that the high energy component has almost 
disappeared. This is expected because the 800 meters of rock overburden highly 
suppresses the intensity of cosmic rays from the atmosphere. After the background is 
subtracted, as seen in the upper-right plot of Fig.~\ref{ambeunderground}, 
the low energy gamma rays from the internal contamination are 
thoroughly removed. The background subtraction is not very effective at the  high energy
muon tail due to the limitation of our statistics. The 4.4 MeV gamma ray line 
from the AmBe source is detected by the detector shown in Fig.~\ref{ambeunderground}. 
A 1.3 MeV energy cut is set on the neutron band to cut off data that is affected 
by random noise at low energy. This 1.3 MeV visible energy is equivalent to 4 MeV
recoil energy by applying the quenching factor measured for EJ301 (NE213) and BC-501A by several groups~\cite{nori, rac, aak} 
including KamLAND~\cite{quenching}. 
Quenching factor is a convolution of the ionization efficiency 
and scintillation efficiency induced by nuclear recoils~\cite{quenchMEI}.  
These two efficiencies are well described by Lindhard's theory~\cite{lindhard} 
and Birks law~\cite{birks}. 
Although the composition of the liquid scintillator we use is a little bit different
from that of KamLAND, both quenching factors induced by nuclear recoils are
dominated by nearly the same ionization efficiency. Therefore, a small difference of less than 5\% in the 
quenching factor between EJ301 (NE213) and  KamLAND's scintillator offers a cross check and confirms our use of the quenching factor.  
Pure neutrons are obtained after appropriate separation
from gamma ray and noise as shown in the bottom left plot of Fig.~\ref{ambeunderground}. 
In order to determine the recoil energy, the quenching
factor is applied to the visible energy of the neutrons, 
as shown in the bottom right plot Fig.~\ref{ambeunderground}.
A GEANT4 based simulation has been performed to compare with 
the measurement data. A good agreement is found for recoil energies 
above 4 MeV. 
\par
The neutrons and gamma rays are not well separated around 4 MeV, as shown in the upper right
plot of Fig. \ref{ambeunderground}. This indicates
that 4 inch thick lead bricks might not be thick enough to block such high energy
gamma rays. The neutron signals are still suffering  
contamination from gamma rays, although there is an improvement over no shielding.
In addition, the low energy range of the neutron band has
an overlap with the gamma ray band. This is
because the pulse shape of low energy
events is easily affected by the attenuation of light in the tube  
and the fluctuation of the pedestal level of the ADC. The sources of the fluctuation are
from the temperature, electronic noise, etc. 
A 23 dB attenuator applied to the signal makes 
the signal-to-noise ratio even worse.
The trigger rate is very high in the low energy
range, which was mainly contributed as internal contamination in the detector.
 It will become a 
challenge if we attempt to measure low energy ($\alpha$,n) neutrons in an underground laboratory. 
We may have to consider sacrificing the sensitivity at low energies 
to improve the data quality.  
\par
\begin{figure}
\includegraphics[angle=0,width=\textwidth]{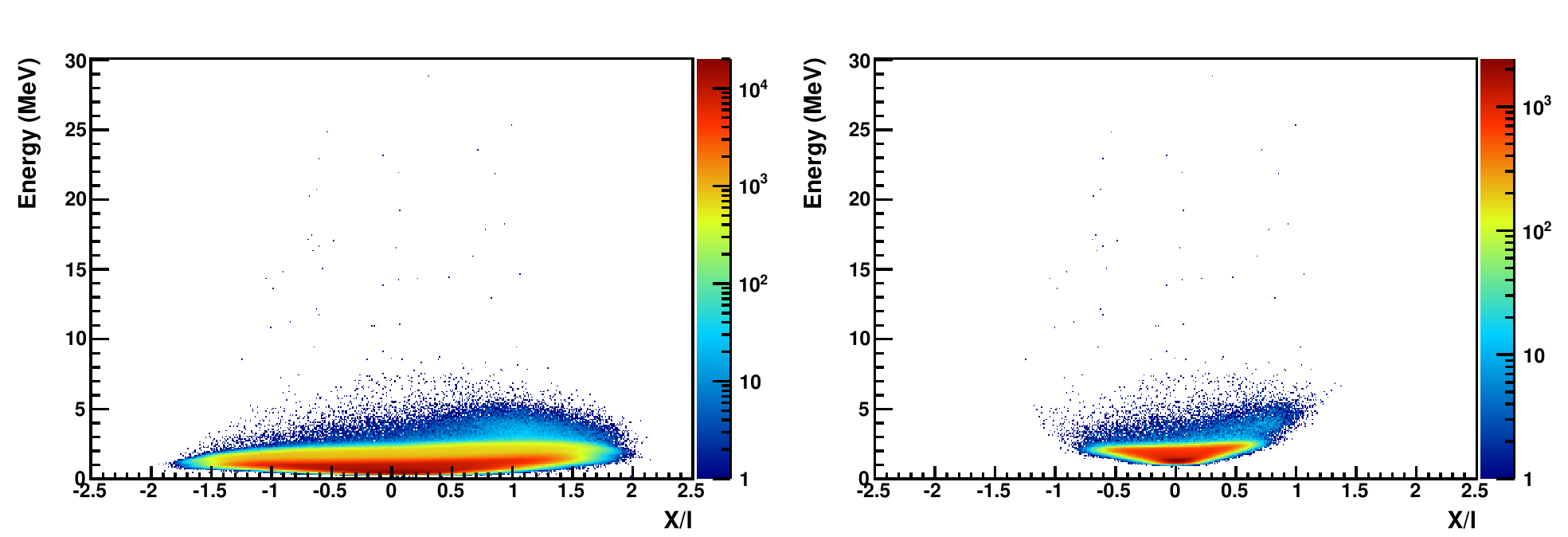}
\includegraphics[angle=0,width=\textwidth]{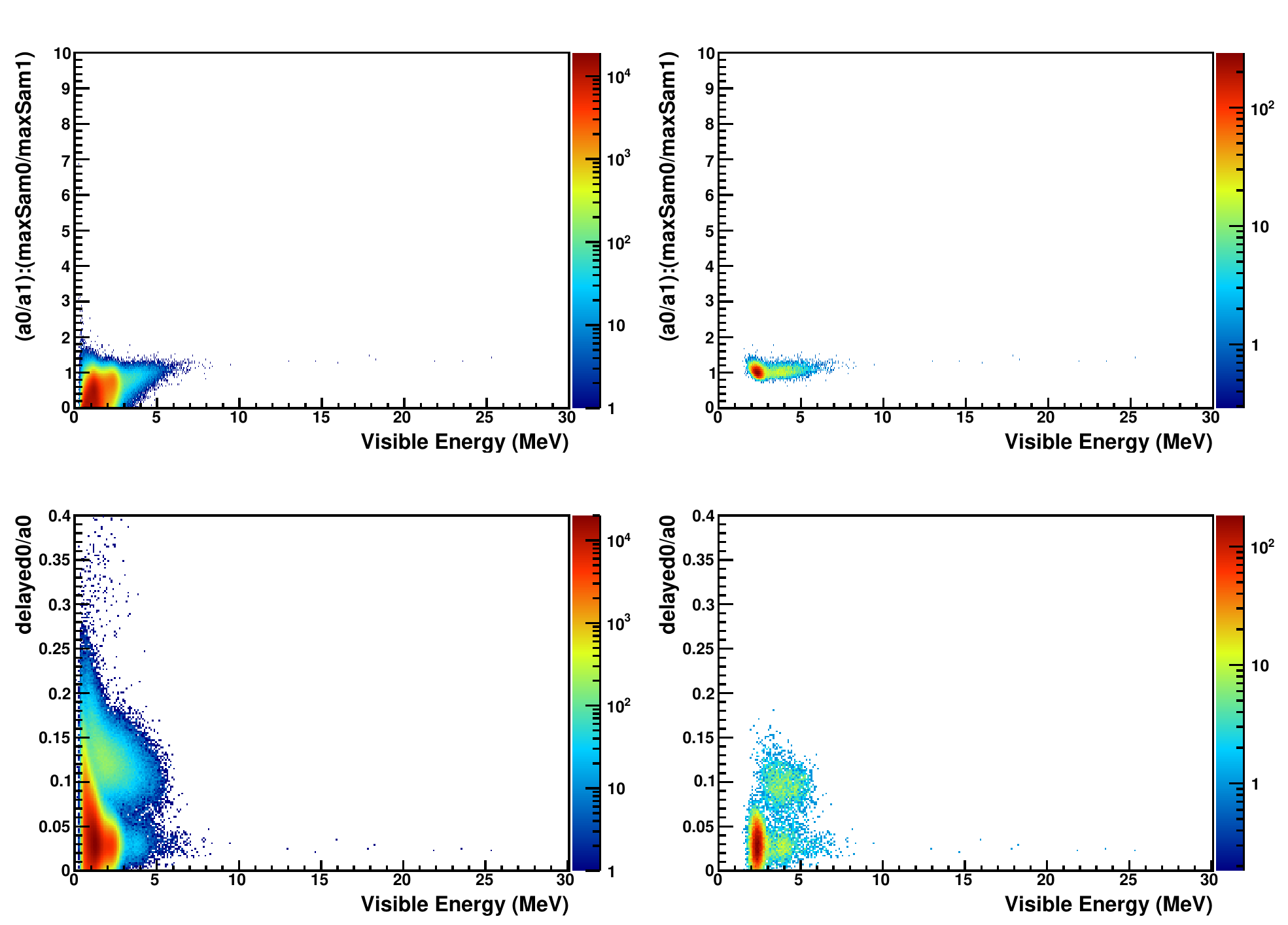}
\caption{\label{th1400} The improvement of data quality
from raising the energy threshold for both PMTs. The left three
plots are the data collection with the conditions:
$ maxSam0>30$, $maxSam1>30$; 
the right three plots are the results with cuts:
$maxSam0>110$, $maxSam1>130$. The bottom four plots apply
a further cut on $X/l>0.5$.   
                }
\end{figure}

A test has been conducted using the same AmBe run, but with a raised
the energy threshold as shown in Fig. \ref{th1400} (maxSam = maximum sample - pedestal) for both PMTs. 
The top two plots show those events with low energies and the 
positions close to one PMT. This causes a weak signal at the other PMT
that is removed by raising the threshold.  
These events are poor quality in terms of their pulse shapes to
at least one of the PMTs. 
Since the value of maxSam 
represents the magnitude of the energy deposition in
the same way as the integrated charge, the ratio of 
$(a0/a1):(maxSam0/maxSam1)$ should be a constant if both
PMTs are triggered by the same event. 

The middle two plots in Fig. \ref{th1400} clearly show how  
low energies are affected by poor quality events and 
how they have been improved by raising the energy threshold.
The ratio, $(a0/a1):(maxSam0/maxSam1)$ is found 
to be a constant ($\sim$1) from 2 MeV up to a few ten MeV.
The comparison of the bottom two plots in Fig. \ref{th1400}
gives more information. First, the phenomenon of 
neutron-gamma superposition in the low energy range is eliminated.
Second, neutrons can be easily identified from the gamma ray band. 
Finally, gamma rays are well characterized by themselves, e.g. the 2.6 MeV
gamma ray line from radioactive decay and 
the 4.4 MeV gamma-ray line from the AmBe source. The 
high energy events are barely affected by raising the threshold.
It is worth pointing out that the energy threshold of the neutron band
is slightly larger than the gamma ray band in the bottom right plot Fig. \ref{th1400}. 
This is because we set the same threshold on the maximum sample
for both neutrons and gamma rays. However, the delayed pulses induced by neutrons is larger than that of gamma rays. Therefore, the 
energy of neutrons represented by the integrated area is
slightly larger than that of the gamma rays.     
\section{Discussion and Conclusion}
Other than protons, high energy neutrons induce recoil deuteron, 
alphas, and $^{12}$C in 
scintillators, which causes distinct bands in the pulse shape
spectrum for small size detectors~\cite{deuterons}. For a large detector, 
these features are also observed at positions
very close to the PMTs. For those recoils occurring away from PMTs, 
the features are washed out due to the complexity of light transmission.   
\par
With a 12 liter liquid scintillation detector, the surface muon minimum 
ionization peak is utilized to calibrate 
the detector with an energy of up to $\sim$20 MeV.
The representation of position using the parameter $X/l$ is found
to be very convenient for interpreting the features of light 
transport in the scintillator. 
Energy independence, in terms of position, is found based 
on a simple mechanism, which makes the energy 
reconstruction much easier. It is also useful to separate neutrons from gamma rays
when a wide range of positions are combined. because of this, we have demonstrated  
the position dependent neutron-gamma separation.  
A pulse shape analysis procedure has been performed to 
distinguish neutrons from  gamma rays.
A new algorithm has been developed for a large liquid scintillation 
detector to directly measure neutrons at a few MeV to a few hundred MeV.   
\par
The direct measurement of fast neutrons, traditionally, can be performed using the time of flight (TOF) technique.
However, the small solid angle coverage dramatically limits its efficiency. 
Because of the extremely low intensity of neutrons, it is almost
impossible to perform such a measurement for the fast neutrons in a deep underground laboratory. 
Rather than using TOF measurements, we provide a new method to directly
measure high energy neutrons with a much better detection efficiency
($\sim$30\% at 10 MeV for the AmBe run) and at a relatively low cost.
Depending on the needs of the underground depth, an array of such neutron detector
modules could be employed to increase the detection efficiency.       
We conclude that a large scintillation detector can be used to measure fast neutrons for 
ultra-low background experiments underground.

\section{Acknowledgement}
The authors wish to thank Yongchen Sun, 
Keenan Thomas, Christina Keller,  
Kareem Kazkaz, and Priscilla Cushman for their invaluable suggestions and help. 
This work was supported in part by NSF PHY-0758120, PHYS-0919278,
PHYS-1242640, DOE grant DE-FG02-10ER46709, 
the Office of Research at the University
of South Dakota and a 2010 research center support by the State of South Dakota.

\end{document}